\documentclass{vienna-conf2019}
\usepackage{graphicx}
\usepackage{hyperref}
\usepackage[]{natbib}  
\usepackage{epstopdf}

\def\BibTeX{{\rm B\kern-.05em{\sc i\kern-.025em b}\kern-.08em
    T\kern-.1667em\lower.7ex\hbox{E}\kern-.125emX}}
\bibpunct{(}{)}{;}{a}{}{,}  
\newcommand{\kms}{{\mathrm{km~s^{-1}}}}

\begin{document}

\TitreGlobal{Stars and their variability observed from space}


\title{The prototype star \boldmath{$\gamma$} Doradus observed by TESS}

\runningtitle{The prototype star $\gamma$ Doradus observed by TESS}

\author{S. Christophe}\address{LESIA, Observatoire de Paris, Universit\'e PSL, CNRS, Sorbonne Universit\'e, Universit\'e de Paris, 5 place Jules Janssen, 92195 Meudon, France}

\author{V. Antoci}\address{Stellar Astrophysics Centre, Department of Physics and Astronomy, Aarhus University, Ny Munkegade 120, DK-8000 Aarhus C, Denmark}

\author{E. Brunsden}\address{Department of Physics, University of York, Heslington, York, YO10 5DD, UK}

\author{R.-M. Ouazzani$^1$}

\author{S.J.A.J. Salmon}\address{STAR Institute, Universit\'e de Li\`ege, All\'ee du 6 Ao\^ut 19, 4000 Li\`ege, Belgium}




\setcounter{page}{237}


\maketitle


\begin{abstract}
$\gamma$ Doradus is the prototype star for the eponymous class of pulsating stars that consists of late A-early F main-sequence stars oscillating in low-frequency gravito-inertial modes. Being among the brightest stars of its kind (V = 4.2), $\gamma$ Dor benefits from a large set of observational data that has been recently completed by high-quality space photometry from the TESS mission. With these new data, we propose to study $\gamma$ Dor as an example of possibilities offered by synergies between multi-technical ground and space-based observations. Here, we present the preliminary results of our investigations.
\end{abstract}



\section{TESS photometry}

\begin{figure}[ht!]
 \centering
 \includegraphics[width=0.39\textwidth,clip]{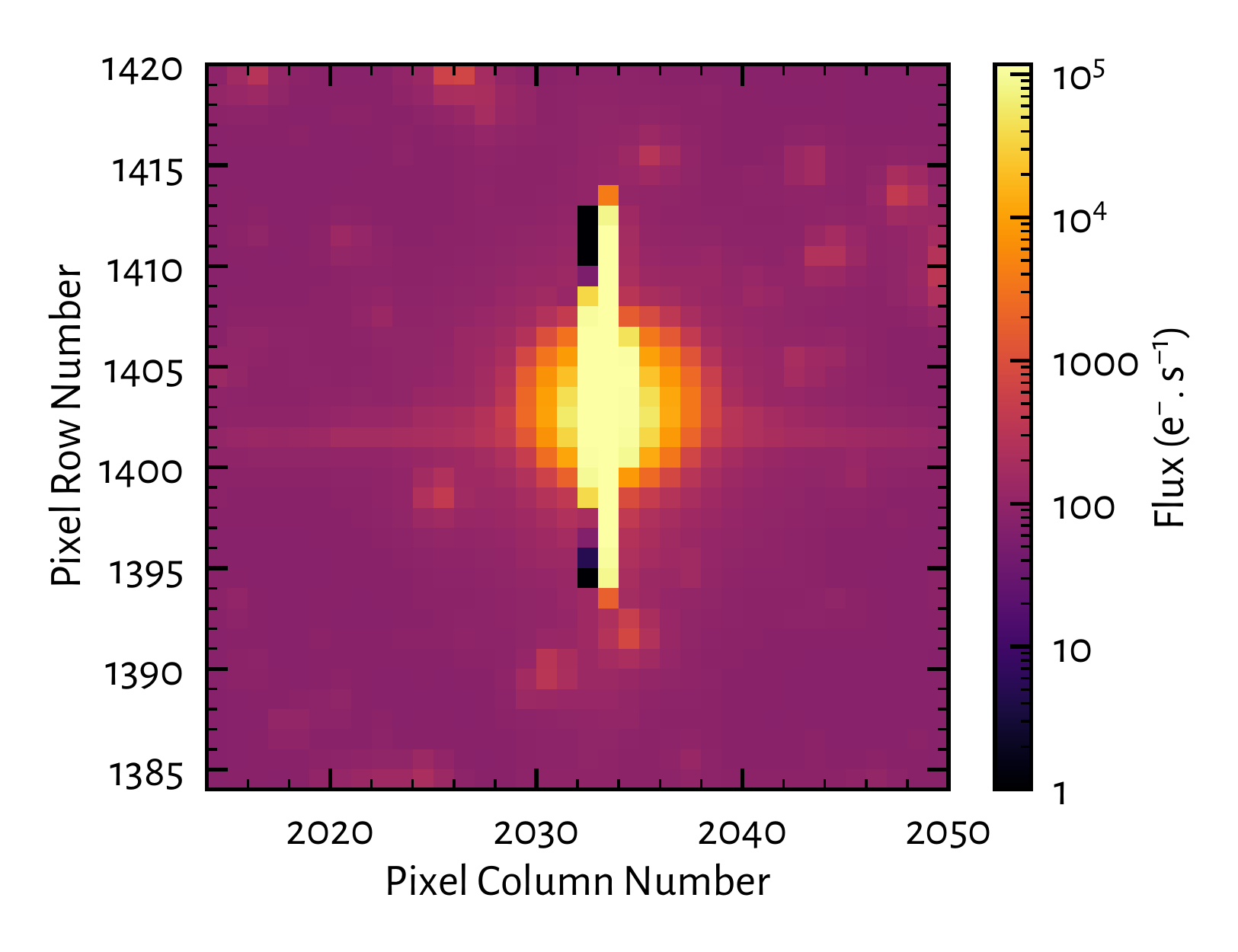}%
 \includegraphics[width=0.61\textwidth,clip]{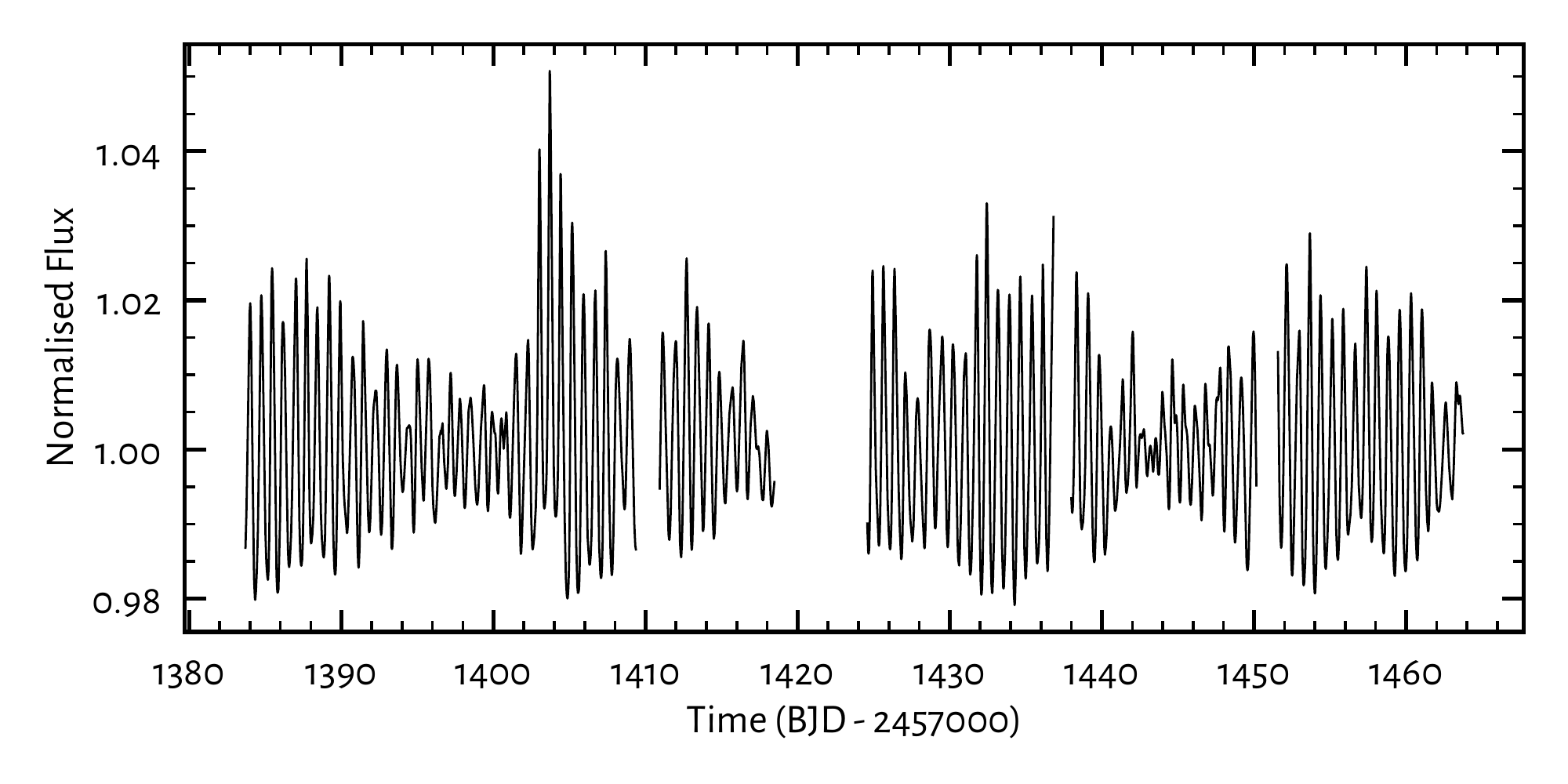}      
  \caption{{\bf Left:} Example of TESS Full Frame Image (FFI) cutout for $\gamma$ Dor. {\bf Right:} Lightcurve of $\gamma$ Dor extracted from TESS FFI cutouts.}
  \label{christophe:fig1}
\end{figure}

TESS observed $\gamma$ Dor during Sectors 3, 4, and 5, representing 80 days of nearly uninterrupted data. Bright as it is, the star saturates the CCDs and leaves bleeding trails along CCD columns. This is shown on the Full Frame Image (FFI) cutout on the left panel of Fig.~\ref{christophe:fig1}. The lightcurve was extracted by performing simple aperture photometry on FFI cutouts with a custom mask chosen to include most of the target flux. Data downlink or scattered light from the Earth or the Moon are responsible for gaps in the data but overall, a duty cycle of 87\% is achevied.

\begin{figure}[ht!]
 \centering
 	\begin{minipage}[c]{0.47\textwidth}
 		\includegraphics[width=\textwidth, trim = 0.6cm 0.6cm 0.6cm 0.6cm, clip]{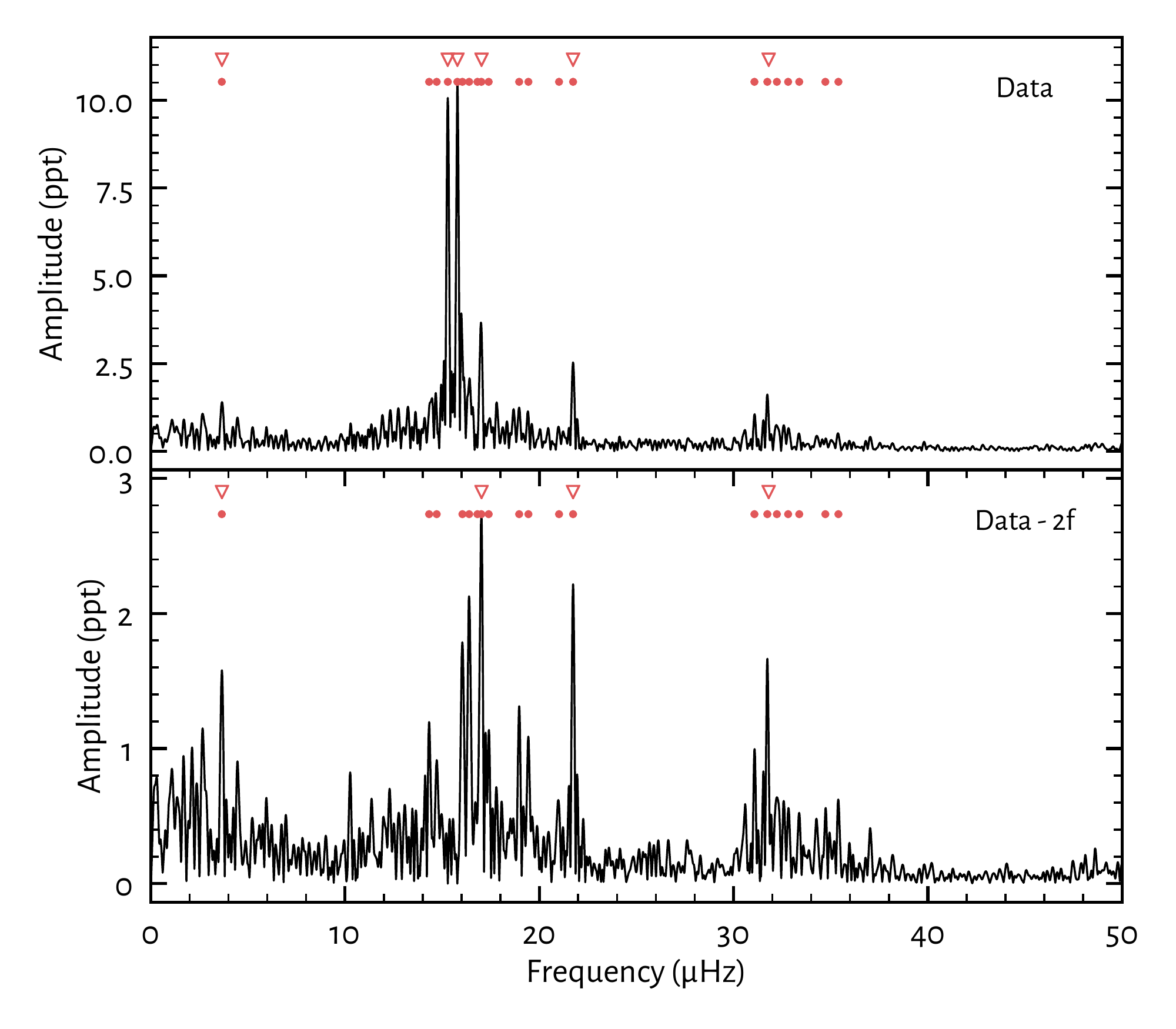}      
	\end{minipage}\hfill
	\begin{minipage}[t]{0.45\textwidth}
  		\caption{Amplitude periodograms before (top) and after pre-whintening (bottom) of the first two frequencies of high amplitude. Red circles represent frequencies extracted from the TESS lightcurve. Upside down triangles indicate frequencies found by \citet{2018MNRAS.475.3813B} from ground-based photometry and spectroscopy.}
 		\label{christophe:fig2}
	\end{minipage}
\end{figure}

\section{Asteroseismology from ground and space}

We carried out the frequency analysis of the TESS lightcurve by iterative prewhitening. We found 21 frequencies with S/N $ > 4$ (see Fig.~\ref{christophe:fig2}). The 6 previously known frequencies from ground-based observations \citep{2018MNRAS.475.3813B} are all confirmed with TESS data. Based on line profile variations, \citet{2018MNRAS.475.3813B} identified the three frequencies at 15.3, 15.8 and 21.7 $\mu$Hz to be prograde gravity modes of $(\ell,m)=(1,-1)$. Assuming the frequency group at $\sim$16 $\mu$Hz is mostly consisting of $(1,-1)$ modes, we applied the method of \citet{2018A&A...618A..47C} to estimate the near-core rotation rate of $\gamma$ Dor. The mode frequencies are compatible with rotation rates in the 9-12~$\mu$Hz range but the limited resolution of the periodogram prevent us from obtaining a more precise value. The second frequency group around $\sim$32~$\mu$Hz can be interpreted as either combination frequencies or $(\ell,m)=(2,-2)$ modes.

\section{Surface rotation}

In order to determine the surface rotation rate, we have need of the luminosity $L$, the effective temperature $T_{\rm eff}$, the projected surface velocity $v\sin i$ and the inclination angle $i$. The luminosity is adopted from Gaia DR2 \citep[$L = 6.99 \pm 0.06~L_\odot$, ][]{2018A&A...616A...1G}. We derive $T_{\rm eff} \approx 7145 \pm 150$~K by averaging the photometric and spectroscopic estimates available in the literature. The projected velocity $v\sin i = 59.5 \pm 3~\kms$ is taken from \citet{2012A&A...542A.116A}. $\gamma$ Dor hosts a debris disc that has been observed with {\it Herschel} and modelled by \citet{2013ApJ...762...52B}, who found $i \approx 70^\circ$. Assuming the rotation axis of $\gamma$ Dor is aligned with its disc axis, the surface rotation rate is estimated to be $8.4 \pm 0.8$ $\mu$Hz. This is close to the range of near-core rotation rates estimated from seismology suggesting a nearly uniform rotation profile.

\begin{acknowledgements}
This research made use of {\sc Lightkurve}, a Python package for Kepler and TESS data analysis \citep{2018ascl.soft12013L}; {\sc Period04} \citep{2005CoAst.146...53L} and {\sc Astroquery} \citep{2019AJ....157...98G}. S.C. acknowledges support from the Programme National de Physique Stellaire (PNPS) of the CNRS/INSU co-funded by CEA and CNES.
\end{acknowledgements}

\bibliographystyle{aa}  
\bibliography{christophe_3p08} 

\end{document}